\documentclass[twocolumn,aps,prb,epsfig]{revtex4}
\usepackage{graphicx}

\newcommand{\nl}{\nonumber \\}

\newcommand{\be}{\begin{equation}}
\newcommand{\ee}{\end{equation}}
\newcommand{\bea}{\begin{eqnarray}}
\newcommand{\eea}{\end{eqnarray}}

\newcommand{\Eq}[1]{Eq.\,(\ref{#1})}

\newcommand{\la}{\langle}
\newcommand{\ra}{\rangle}
\newcommand{\dg}{\dagger}

\newcommand{\ti}{\tilde}

\begin{document}
\draft

\title{Quantum computation with coupled-quantum-dots embedded in optical microcavities}

\author{Xin-Qi Li$^{1}$ and YiJing Yan$^{2}$}

\address{$^{1}$National Laboratory for Superlattices and Microstructures,
         Institute of Semiconductors,
         Chinese Academy of Sciences, P.~O.~Box 912, Beijing 100083, China \\
         $^{2}$Department of Chemistry, Hong KOng University of Science and Technology,
         Kowloon, Hong Kong }

\date{January 4, 2002; to appear in Phys. Rev. B; code\#:BN8032}

\begin{abstract}
Based on an idea that spatial separation of charge states can
enhance quantum coherence, we propose a scheme for quantum
computation with quantum bit (qubit) constructed from two coupled
quantum dots. Quantum information is stored in electron-hole pair
state with the electron and hole locating in different dots, which
enables the qubit state being very long-lived. Universal quantum
gates involving any pair of qubits are realized by coupling the
quantum dots through cavity photon which is a hopeful candidate to
transfer long-range information. Operation analysis is carried out
by estimating the gate time versus the decoherence time.
\end{abstract}

\vspace{3ex}

\maketitle


To build a practical quantum computer is a challenging task,
since the computational quantum objects (the qubits)
must be suffficiently isolated from the dissipative environment,
precisely and conditionally manipulated,
efficiently read-out and initialized, and most importantly scalable
\cite{Div95255,Eke96733,Ste98117}.
To date, a variety of quantum computation (QC) schemes based on
some unique systems have been proposed
\cite{Tur954710,Cir954091,Ger97350,Kan98133,Ave98659,Los98120,Kni0146,Pla991967,Ber005912,Pri98141,Kit9707}.
Although in a number of systems the proof-of-principle has been
convincingly demonstrated, great challenges exist in achieving a
useful quantum computer. No single system has emerged as a clear
leading candidate, each having its merits and drawbacks with
respect to the requirement of scalability and fault-tolerance. In
regard to the scalability, the solid-state implementations should
represent one of the most promising directions.


A representative possibility for solid-state QC is the state-in-art technology
based on quantum dots (QDs).
In quantum dots, the discrete electronic charge states or spin states can be exploited
to encode quantum information, and the transfer of information between qubits
can be mediated, for example, by electron-electron Coulomb interaction
\cite{Bar95,San99,Tan00,Li01,Bio00,Qui99,Sham01},
or by optical cavity photon or vibrational phonon as a data-bus
\cite{She99,Ima99,Bru00,Bro01}.
In particular, impressive experimental progress on coherently operating and entangling
charge and spin states in QDs have been reported very recently \cite{Sham00,Haw01,Gup01}.

The main drawback to encode quantum information in charge states of quantum dots
is the severe decoherence. To overcome it, a possible way is to apply relatively
strong laser pulses to perform {\it ultrafast} operations
\cite{Bar95,Bio00,Qui99,Sham00,Haw01,Gup01}.
In our recent work \cite{Li01}, a scheme to reduce the decoherence of
charge states in quantum dots was proposed to build up a single qubit
from two coupled QDs. We showed that the spatial separation of the logic states
can efficiently reduce the qubit decoherence.
Nevertheless, several shortcomings exist there and in some of the aforementioned
QC schemes based on QDs:
(i) In each qubit (two coupled QDs), one and only one excess electron is required
in the conduction band. This is a challenging task within current technology.
(ii) The intersubband transition with THz lasers is currently not a mature technology.
(iii) The coupling between qubits is mediated by Coulomb interactions,
which makes it very difficult to perform conditional gate operation
between any pair of qubits.

In this paper, still based on the idea by constructing a single qubit
from two coupled QDs to reduce decoherence, we propose an alternative scheme
to remove all of these shortcomings.
In this newly proposed scheme, no excess electron is required in the qubit, and
quantum information is stored in electron-hole pair state.
An all-optical approach is suggested for transition between valence and conduction band states
that only require the well-known available laser sources.
Viewing the advantages of the cavity QED effect, the cavity photon is believed to be
an ideal candidate to transfer information, with which conditional quantum gate
operation can be performed between any pair of qubits.
In the proposed structure, quantum measurement of the qubit states can be achieved
also optically through the quantum state holography.
Specifically, the amplitude and phase information of the qubit state
(i.e. the electron-hole pair state), can be extracted
through mixing it with a reference state generated in the same system
by an additional delayed laser pulse and detecting the total time- and frequency-
integrated fluorescence as a function of the delay time \cite{Lei95,Wei9899}.



The physical system we are concerned with for quantum computation is similar
as that proposed by Imamogl\={u} {\it et al} \cite{She99,Ima99}, i.e.,
many QDs are located in an optical microcavity.
Both the QDs and cavity are three-dimensionally (3D) confined; however, the cavity has
a size and thus the fundamental wavelength much larger than the individual QD.
In our structure, we suggest to use two weakly coupled QDs to construct a qubit
as shown in Fig.\ 1.
We assume a relatively large distance between neighboring qubits
such that the qubits can be selectively addressed by lasers, and the Coulomb
correlations between them can be neglected.
As have mentioned above, in our structure no excess single
electron is required in the conduction band. The quantum
information is stored in electron-hole pair state: the qubit logic
states $|\ti{0}\ra$ and $|\ti{1}\ra$ correspond to the ground
state and an electron-hole pair state, respectively. (Here the
symbol {\it tilde} is used to distinguish the notation of logic
states from the cavity states ). A key idea of this work is to
create the electron-hole pair state with electron and hole
locating largely in different dots of the qubit, which is expected
to have much longer lifetime than its counterpart in a single dot.
In the following, we detail two means of qubit transition as shown
in Fig.\ 1: one involves the cavity-photon participation; another
does not. As will be shown later, the cavity-photon can play a
data-bus role to couple any pair of qubits.

\begin{figure}\label{Fig1}
\begin{center}
\centerline{\includegraphics [scale=0.25] {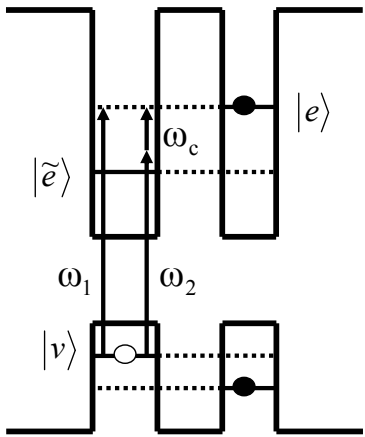}} \caption{
Schematic diagram for a qubit constructed from two coupled quantum
dots with different sizes. The plotted states are resulted from
the HOMO and LUMO of the individual quantum dots. The ground state
denoted by $|v\ra$ is used for the qubit logic state $|\ti{0}\ra$,
and the excited state $|e\ra$ for the logic $|\ti{1}\ra$.
$|\ti{e}\ra$ plays a role of intermediate state with virtual
occupation in the qubit operation with two photon participation.
The optical coupling between states are due to the classical
lasers with frequencies $\omega_1$ and $\omega_2$, and cavity
photon with frequency $\omega_c$. }
\end{center}
\end{figure}

The relevant electronic states for optical transition are shown in Fig.\ 1
by $|e\ra$, $|\ti{e}\ra$ and $|v\ra$, respectively, which are resulted
by accounting for weak coupling
from the HOMO and LUMO states of the two individual QDs with different sizes.
Due to the confinement, we assume no intermediate levels between the two lowest
conduction band states and between the highest valence band states.
In general, in the absence of magnetic field, both the HOMO and LUMO levels
are spin degenerate.
However, in the present proposal, we exploit the charge states rather than
the spin states to store quantum information, thus the superposition and
even the decoherence of spin states are irrelevant to the logic states.
For this reason, the spin index is omitted in the state notations.
We first consider the qubit operation involving no cavity photon
participation. In this case, by turning on a travelling-wave laser
field with frequency $\omega_1$ on resonance with the energy
difference between $|e\ra$ and $|v\ra$, the Rabi transition takes
place under the interaction \bea\label{HI1} H_{I}^{(1)}
     = \Omega_1 \left[ |e\ra \la v| e^{i\phi}
       + \mbox{H.c.}  \right] .
\eea Here $\Omega_1$ is the Rabi frequency, and $\phi$ is the
laser phase. With the use of this interaction, arbitrary
single-qubit operations can be performed.

Next, consider the cavity-photon involved transition.
This is an essential ingredient to
realize the two-qubit gate, in which the cavity photon plays a role
of data-bus to control the two-qubit state evolution.
Switching on a laser action with frequency $\omega_2=E_{e}-E_v-\omega_c$,
a resonant transition from $|v\ra$ to $|e\ra$ takes place
by involving two photons, namely, the $\omega_2$ laser photon and the $\omega_c$
cavity photon. A simple perturbation theory gives rise to
\bea\label{HI2}
H_{I}^{(2)}
     = \Omega_{\mbox{eff}} \left[ |e\ra \la v|a e^{i\phi}
       + \mbox{H.c.} \right] ,
\eea
with $\Omega_{\mbox{eff}}=\Omega_2\Omega_c/\delta$.
Here $\Omega_2$ is the optical coupling strength between $|\ti{e}\ra$ and $|v\ra$
associating with the laser field,
and $\Omega_c$ is the coupling strength between $|e\ra$ and $|\ti{e}\ra$
due to the cavity field.
$\delta$ is the detuning between the laser frequency and the transition energy
from $|v\ra$ and $|e\ra$, i.e., $\delta=\omega_2-(E_{\ti{e}}-E_v)$.
Note that this second-order process is mediated via the intermediate state $|\ti{e}\ra$.
However, due to the off-resonance of the laser frequency with the transition
$|v\ra\rightarrow |\ti{e}\ra$, there is no real occupation on state $|\ti{e}\ra$,
accordingly its relatively strong decoherence resulting from its radiative
recombination with the intra-dot hole $|v\ra$ is avoided.
In the latter part of this paper, we will show that
owing to the spatial separation of $|e\ra$ from the states in the larger dot,
the coherence of qubit state $|e\ra$ can be essentially improved.


To realize the conditional two-bit gate such as the control NOT (CNOT),
typical methods include the Cirac-Zoller (CZ) protocol \cite{Cir954091},
or the pulse technique developed in the spin QC model \cite{Los98120}.
In the proposals based on QDs in cavity, both of these two gating techniques
have been employed \cite{She99,Ima99,Bro01}.
Very recently, an improved gating technique for the ion-trap QC was developed
where only two electronic states are required, and the third auxiliary state
in the CZ protocol is not needed \cite{Chi01}.
In the following, we employ this technique in our scheme with certain modification
due to the only {\it red-band pulse} in our case.
To make the description more transparent, we introduce the
following notations: the states of the control qubit (the $j$th
one) and the target qubit (the $k$th one) together with the cavity
photon are denoted as $\{|\ti{a}_j\ti{b}_k\ra|p\ra \equiv
|\ti{a}_j\ra|\ti{b}_k;p\ra: a,b,p=0,1 \}$, where 0 and 1
correspond to either the qubit state (i.e. logic
$|\ti{0}\ra\equiv|v\ra$ and $|\ti{1}\ra\equiv |e\ra$), or the
cavity field state with zero and one photon.
Below we outline how to realize the CNOT gate.

{\bf (i)}
First, swap the control qubit state to the cavity photon state:
\bea\label{Rmj}
R_j(\pi,\phi) \left[ \begin{array}{c}
                |\ti{0}_j\ti{0}_k\ra |0\ra   \\
                |\ti{1}_j\ti{0}_k\ra |0\ra   \\
                |\ti{0}_j\ti{1}_k\ra |0\ra   \\
                |\ti{1}_j\ti{1}_k\ra |0\ra   \\
           \end{array} \right]
  = |\ti{0}_j\ra    \left[ \begin{array}{r}
                         |\ti{0}_k; 0\ra   \\
             ie^{-i\phi} |\ti{0}_k; 1\ra   \\
                         |\ti{1}_k; 0\ra   \\
             ie^{-i\phi} |\ti{1}_k; 1\ra   \\
                     \end{array} \right]  .
\eea
Hereafter the evolution operator $R_{j(k)}(\theta,\phi)$
is determined by the interaction Hamiltonian  $H_{I}^{(2)}$ in terms of
$R(\theta,\phi)=\mbox{exp}\left[
   i\frac{\theta}{2} \left( |e\ra \la v|a e^{i\phi}
   + \mbox{H.c.} \right)\right]$,
where $\theta=2\Omega_{\mbox{eff}} T$
with $T$ the duration time of the laser pulse.

{\bf (ii)}
Now, the cavity photon can play a role of control qubit, which controls the evolution of
the target qubit.
A series of pulse operations on the $k$th qubit by involving the participation
of the cavity photon yield
\bea\label{Gk}
G_k \left[ \begin{array}{c}
                         |\ti{0}_k; 0\ra   \\
                         |\ti{0}_k; 1\ra   \\
                         |\ti{1}_k; 0\ra   \\
                         |\ti{1}_k; 1\ra   \\
           \end{array} \right]
=   \left[ \begin{array}{r}
                        \delta   |\ti{0}_k; 0\ra   \\
                       -\delta^* |\ti{1}_k; 1\ra   \\
                        \delta   |\ti{1}_k; 0\ra   \\
                       -\delta^* |\ti{0}_k; 1\ra   \\
           \end{array} \right] ,
\eea
where $\delta=e^{i\phi_0}$, and $\phi_0=\pi/2\sqrt{2}$.
We see that the cavity photon has played a control role in the conditional evolution
of the $k$th qubit.
In \Eq{Gk}, $G_k$ constitutes a series of operations on the $k$th qubit,
$G_k \equiv H_k[P_kZ_k(\phi_0)]H_k$.
Here $H_k$ and $Z_k(\phi_0)$ are the single-qubit Hadamard
and phase transformations
\bea  \label{HZ}
H_k = \frac{1}{\sqrt{2}} \left[ \begin{array}{cc}
                             -1  &  1    \\
                              1  &  1
                       \end{array} \right]  , ~~~~~
Z_k(\phi_0)= \left[ \begin{array}{cc}
               e^{i\phi_0}  &  0  \\
               0   & e^{-i\phi_0}
           \end{array} \right]  .
\eea
In the subspace
$\{|\ti{0}_k;0\ra,|\ti{0}_k;1\ra,|\ti{1}_k;0\ra,|\ti{1}_k;1\ra \}$,
the operator $P_k$ has a diagonal form
\bea \label{Pk}
  P_k &\equiv& R_k(-\pi/2,0)R_k(\sqrt{2}\pi,-\pi/2)R_k(\pi/2,0)  \nl
      & = &    \mbox{diag}(1,e^{-i\pi/\sqrt{2}},e^{i\pi/\sqrt{2}},-1) .
\eea

{\bf {(iii)}}
Finally, the cavity photon state is swapped back to the qubit state by performing
operation $R_j(\pi,\phi)|\ti{0}_j;1\ra = ie^{i\phi}|\ti{1}_j;0\ra$, on the $j$th-qubit.
After a phase gate $Z_j(-\phi_0)$ on the $j$th-qubit, the standard CNOT gate
is realized
\bea\label{CNOT}
U_{jk}
    \left[ \begin{array}{c}
                |\ti{0}_j\ti{0}_k\ra |0\ra   \\
                |\ti{1}_j\ti{0}_k\ra |0\ra   \\
                |\ti{0}_j\ti{1}_k\ra |0\ra   \\
                |\ti{1}_j\ti{1}_k\ra |0\ra   \\
           \end{array} \right]
=  \left[ \begin{array}{c}
                |\ti{0}_j\ti{0}_k\ra |0\ra   \\
                |\ti{1}_j\ti{1}_k\ra |0\ra   \\
                |\ti{0}_j\ti{1}_k\ra |0\ra   \\
                |\ti{1}_j\ti{0}_k\ra |0\ra   \\
           \end{array} \right]  ,
\eea
where $U_{jk}=Z_j(-\phi_0)R_j(\pi,\phi)G_kR_j(\pi,\phi)$.


In the remainder part of this paper, we present an analysis for the QC operation.
In Ref.\ \onlinecite{Li01}, based on a model GaAs system and disk geometry for the QDs,
we have demonstrated by detailed numerical calculations
that the ratio $\rho=\tau_d/\tau_G$ can be enhanced
remarkably by the spatial separation of the qubit states,
where $\tau_d$ and $\tau_G$ are the qubit decoherence and gating time, respectively.
In what follows we provide an alternative way to understand this issue in general,
not specifying the concrete QD material and geometry.

In the approximation of two-level model, $|e\ra$ and $|\ti{e}\ra$ come from
the coupling of two isolated dot states
$|d\ra$ and $|\ti{d}\ra$ with coupling strength $t$ and energy separation
$\Delta=E_{d}-E_{\ti{d}}$. (For the highest two valence band state, similar
treatment can be done).
As a result, the eigenstates $|e\ra$ and $|\ti{e}\ra$ have eigenenergies
$E_{\pm}=\frac{1}{2}[(E_d+E_{\ti{d}})\pm\sqrt{\Delta^2+4t^2}]$, and wavefunctions
\bea\label{WFee}
|e\ra &=& \sqrt{1-\gamma}|d\ra+\sqrt{\gamma}|\ti{d}\ra   \nl
|\ti{e}\ra &=& \sqrt{1-\gamma}|\ti{d}\ra-\sqrt{\gamma}|d\ra  ,
\eea
where $\gamma=t^2/(\Delta^2+t^2)$.
With this state nature in mind, we below estimate various decohence time and
operation time in order.

The decoherence time of the qubit state is characterized by
the relaxation time of $|e\ra$.
In our structure, the main intrinsic decoherence mechanisms come from the radiative
relaxation and electron-phonon scattering.
For both mechanisms, the relaxation rate can be expressed on the basis of Fermi
golden rule as
$ W^{(j)} = \frac{2\pi}{\hbar}\sum_{\bf{q}}|M^{(j)}_{fe}(q)|^2
                          \delta(E_e-E_f-\hbar\omega_q ) $ ,
where $\omega_q$ is the emitted photon (phonon) frequency,
and $M^{(j)}_{fe}$ is the perturbative matrix element
$M^{(j)}_{fe}(q)=\la f|H^{(j)}_{\mbox{ep}}(q)|e\ra$.
Here the index $j=1,2$ and 3, together with the final state $|f\ra$,
denote three relaxation channels, namely,
the spontaneous radiation from $|e\ra$ to $|v\ra$ and $|\ti{e}\ra$,
and the phonon-scattering induced relaxation from $|e\ra$ to $|\ti{e}\ra$.
Due to the spatial separate nature of the electronic states shown in \Eq{WFee},
we roughly estimate that the relaxation rate of each channel would be
reduced by a factor $\gamma$,
in comparison with the relaxation rate in a single dot.
As a consequence, the decoherence time can be considerably enhanced by
$\tau_d\simeq\ti{\tau}_d/\gamma$,
where $\ti{\tau}_d$ is the intra-dot decoherence time.

The operation time is limited by the optical coupling
between $|e\ra$ and $|v\ra$ via the external laser field,
and between $|e\ra$ and $|\ti{e}\ra$ via the cavity photon.
For both cases, the coupling strengths can be expressed in terms of
$\Omega_{1(c)}=\la e|H_I^{(1,c)}|v(\ti{e})\ra$.
Similarly as above, due to the spatial separation of sate $|e\ra$
from $|v\ra$ and $|\ti{e}\ra$ as shown in \Eq{WFee}, $\Omega_1$
and $\Omega_c$ will be reduced approximately by a factor
$\sqrt{\gamma}$ in comparison with the corresponding intra-dot
coupling strengths. From \Eq{HI1} and (2), the logic state
flipping time (between $|\ti{0}\ra)$ and $|\ti{1}\ra)$) is
$\pi/\Omega_1$ or $\pi/\Omega_{\mbox{eff}}$, corresponding to the
cavity photon involved or non-involved transition. As a
consequence, the gate ratio $\rho=\tau_d/\tau_G$ will be enhanced
by a factor $\sim 1/\sqrt{\gamma}$ due to the spatial separation
of the qubit states. Note that $\gamma=t^2/(\Delta^2+t^2)$, which
can be a considerably small factor by reducing $t$ and increasing
the energy-level separation $\Delta$. Similar conclusion has been
quantitatively demonstrated by numerical calculation in
Ref.\ \onlinecite{Li01}.

We now parameterize the gate operation and decoherence times.
Generally, consider each qubit consisting of two weekly coupled
quantum dots with coupling strength between $|e\ra$ and
$|\ti{e}\ra$ (see Fig.\ 1) as, for example, $t=0.01$ meV, and
energy difference $\Delta=E_e-E_{\ti{e}}=10$ meV due to the
distinct dot sizes. With these parameters, the {\it spatial
separation factor} $\gamma=10^{-6}$.
Concerning the optical coupling with the electronic states, for
the intra-dot interband transition due to the laser pulse, we
assume a coupling strength $\Omega_2=0.1$ meV; and for the
intra-dot state coupling with the cavity photon, a typical value
of $\ti{\Omega}_c=300$ MHz is adopted here \cite{She99,Ima99}. To
avoid a real occupation on the state $|\ti{e}\ra$, a detuning
$\delta=1$ meV is assumed between the laser frequency and the
energy difference between $|\ti{e}\ra$ and $|v\ra$.
As an approximate estimate, for the cavity-photon involved
transition from $|v\ra$ to $|e\ra$, the effective Rabi frequency
$\Omega_{\mbox{eff}}=\Omega_2\Omega_c/\delta\simeq 30$ KHz; and
for the same transition in the absence of cavity photon, the Rabi
frequency $\Omega_1\sim\sqrt{\gamma}\Omega_2=10^{-4}$ meV. As a
consequence, for single qubit operations, the operation time is of
the order of hundreds nanosecond, whereas for two-qubit
conditional operations, the characteristic time is determined by
the Rabi frequency $\Omega_{\mbox{eff}}$ in terms of
$\tau_G=\pi/\Omega_{\mbox{eff}}\simeq 10^{-4}$ sec.

For the decoherence time, we note that with current technology the
quantum dot is available with energy level spacing
larger than 10 meV, thus we assume no other electronic states
between $|e\ra$ and $|\ti{e}\ra$ in our structure. As
a result, the intrinsic decoherence channels would be the
radiative relaxation and electron-phonon scattering from $|e\ra$
to $|\ti{e}\ra$ and $|v\ra$.
For the spontaneous emission, if the quantum dot has certain
geometric symmetry, it has been shown that the so called dark
states can have radiative lifetime longer than microsecond
\cite{Efr96}. Moreover, due to the CQED effect resulting from
three dimensional cavity with high finesse, the spontaneous
emission lifetime can be further suppressed. With these
considerations, the radiative lifetime of the intra-dot conduction
band state (e.g. radiative transition from $|\ti{e}\ra$ and
$|v\ra$) can be longer than tens to hundreds of microsecond
\cite{Efr96}.
For the decoherence due to interaction with phonons, the energy
level spacing of 10 meV between $|e\ra$ and $|\ti{e}\ra$ can be
much different from the LO phonon energy, thus the phonon induced
relaxation from $|e\ra$ to $|\ti{e}\ra$ is determined by the
acoustic phonon scattering. Following Bockelmann {\it et al}
\cite{Boc90,Ben91}, the electron acoustic-phonon scattering rate
will decrease rapidly with increasing the electronic level
spacing. For example, in Ref.\ \onlinecite{Ima99}, acoustic phonon
scattering time of $\sim 150 ~\mu$s has been carried out for a
similar energy level spacing (i.e., for 12.25 meV).
Therefore, as an order of magnitude estimate, if we adopt an
intra-dot relaxation time ($\ti{\tau}_d$) of tens of microsecond,
the qubit decoherence time can be as long as tens of second
(note that $\tau_d=\ti{\tau}_d/\gamma \sim 10^6 \times \ti{\tau}_d$),
owing to the spatial separation of the qubit states.
Within this time scale, the single bit rotation can be performed
as high as $10^8$ times, and the two-bit CNOT gate can
be performed about $10^4$ times.

Other sources to decoherence include, e.g., the inhomogeneity of quantum dots
and loss of cavity photons. Since the qubit is selectively addressed by
lasers, the former might be overcome by individually tuning the laser
frequency with the qubit states.
The central challenge to realize the proposed QC scheme is the development
of few-mode THz cavities with extremely low loss. An attractive candidate
is the dielectric cavities, which is currently an intensive research field.


\begin{figure}\label{Fig2}
\begin{center}
\centerline{\includegraphics [scale=0.25] {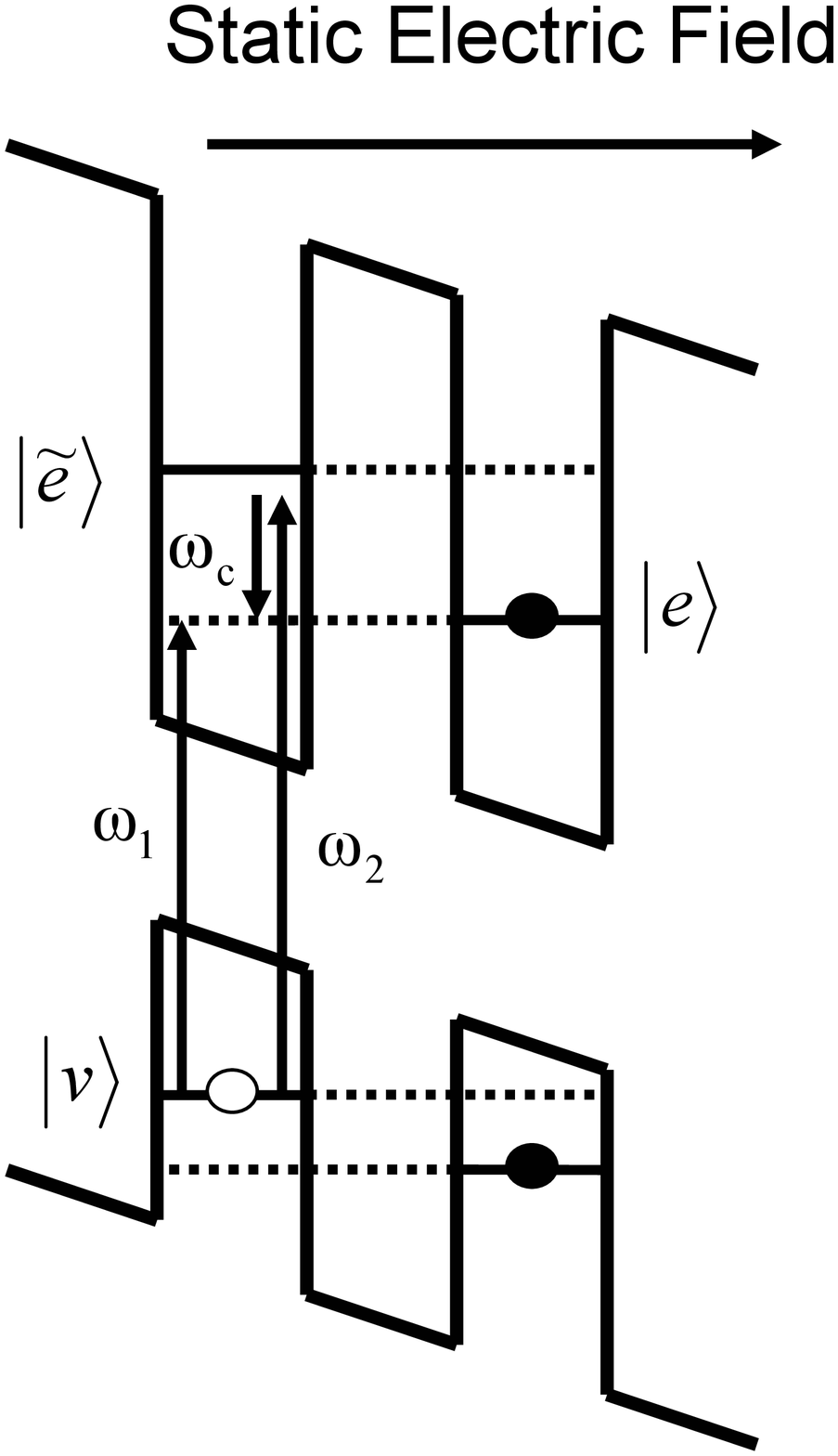}}
\caption{An alternative configuration for qubit construction from
two coupled identical QDs in the presence of external electric
field. This structure is expected to be able to suppress phonon
scattering from $|e\ra$ to $|\ti{e}\ra$ in low temperature regime.
The price paid here is the use of external electric field which
might cause additional dephasing.  }
\end{center}
\end{figure}

In the above discussion for decoherence, we have followed the conventional
approach to neglect the effect of LO phonon scattering due to the large
energy difference between the LO phonon and the electronic level spacing.
This treatment is reasonable if the LO phonon has infinite long lifetime.
It seems however questionable if one takes into account the finite lifetime
of the LO phonons. Our calculations showed that the confined
LO phonons in quantum dots have similar lifetime as in bulk materials,
with the order of magnitude of picosecond \cite{Li98}.
As a consequence, the LO phonon can induce electron relaxation
even under the off-resonance condition \cite{Li99}.
In the weak coupling limit as discussed here due to the spatial separation of
qubit states, it can be shown \cite{Li99} that the LO phonon induced relaxation time
is proportional to $g^{-2}$, with $g$ being the coupling
strength between electron and LO phonons.
Since this effect is still a topic in debate in the context of phonon
bottleneck in quantum dots, we are not quite sure
whether it is a severe decoherence resource in the above proposed QC scheme.

To further improve the phonon scattering induced decoherence, a slightly modified
qubit structure can be designed as follows.
Similar as shown in Fig.\ 1, each qubit still constitutes two quantum dots.
We suggest here to use two identical quantum dots, and to apply constantly
a static electric field that results in an energy level structure
as depicted in Fig.\ 2.
In this qubit structure, the information is also stored in the states
$|v\ra$ and $|e\ra$, and $|\ti{e}\ra$ plays a role of mediating transition
with only virtual occupation on it.
The gate operations based on this structure can be performed similarly
as in the previous one, only noticing that the
cavity-photon involved interaction Hamiltonian is now in the form of
$ H_{I}^{(2)} = \Omega_{\mbox{eff}} \left[ |e\ra \la v|a^{\dg} e^{i\phi}
       + \mbox{H.c.} \right] $, instead of \Eq{HI2}.
As a result, the swap operation corresponds to generating
a cavity photon via the transition from $|v\ra$ to $|e\ra$,
and annihilating a cavity photon vice versa.
The main advantage of this scheme is that the phonon scattering
from $|e\ra$ to $|\ti{e}\ra$ can be almost completely suppressed
in the low temperature limit. In particular, there would be no
LO phonon excitations.
Another merit is that the electric field can be conveniently used to
tune the level spacing between $|e\ra$ and $|\ti{e}\ra$
in near resonance with the cavity photon energy.
The price paid here is the constant presence of an external electric field,
whose thermal fluctuations (the Johnson noise) may cause additional dephasing.
Fortunately, in our scheme the electric field is not varied to perform the
logic operations. Thus the electrodes which generate the electric field
can be connected to a superconducting ground, which can remove the thermal
fluctuations since there is no dissipation.


In summary, we proposed a scheme based on coupled QDs embedded in
optical microcavity to implement quantum computation. The proposed
qubit constructed from two weakly coupled QDs is expected to have
long decoherence time due to the spatial separation of the logic
states. The recent progress of gating technique in ion-trap QC
enables us to realize the universal quantum gate in our structure
based on certain simple electronic state configuration, namely,
the LUMO and HOMO states. From the consideration on the tradeoff
between phonon scattering and fluctuations of electrostatic field,
we suggested two possible qubit configurations for practical
choice.
The most challenging aspect in the proposed QC scheme is to locate QDs in
optical cavity {\it with high finesse}. Modification to the proposed gating
scheme is possible by using the cavity state only as a virtual state, which
can in certain sense relax the requirement to the cavity finesse.

\section*{Acknowledgments}
    Support from the special grant of CAS (Chinese Academy of Sciences) to
distinguished young researchers is gratefully acknowledged.


\begin{references}
\bibitem{Div95255}
D.P. DiVincenzo, Science {\bf 269}, 255 (1995).
\bibitem{Eke96733}
A. Ekert and R. Josza, Rev. Mod. Phys. {\bf 68}, 733 (1996).
\bibitem{Ste98117}
A.M. Steane,
Rep. Prog. Phys. {\bf 61}, 117 (1998).
\bibitem{Tur954710}
Q.A. Turchette, C.J. Hood, W. Lange, H. Mabuchi, and H.J. Kimble,
Phys. Rev. Lett. {\bf 75}, 4710 (1995).
\bibitem{Cir954091}
J.I. Cirac and P. Zoller, Phys. Rev. Lett. {\bf 74}, 4091 (1995).
\bibitem{Ger97350}
D.G. Cory, A.F. Fahmy, and T.F. Havel, Proc. Nat. Acad. Sci. USA {\bf 94}, 1634 (1997);
N.A. Gershenfeld and I.L. Chuang, Science {\bf 275}, 350 (1997);
\bibitem{Kan98133}
B.E. Kane, Nature {\bf 393}, 133 (1998).
\bibitem{Ave98659}
D.V. Averin, Solid State Commun. {\bf 105}, 659 (1998);
Y. Makhlin, G. Sch\"on, and A. Shnirman, Nature {\bf 398}, 305 (1999);
L.B. Ioffe, V.B. Geshkenbein, M.V. Feigelman, A.L. Fauchere,
and G. Blatter, Nature {\bf 398}, 679 (1999);
J.E. Mooij, T.P. Orlando, L. Levitov, L. Tian, C.H. van der Wal, and S. Lloyd,
Science {\bf 285}, 1036 (1999).
\bibitem{Los98120}
D. Loss and D.P. DiVincenzo, Phys. Rev. A {\bf 57}, 120 (1998).
\bibitem{Kni0146}
E. Knill, R. Laflamme, and G. Milburn, Nature {\bf 409}, 46 (2001)  
\bibitem{Pla991967}
P.M. Platzman and M.I. Dykman, Science {\bf 284}, 1967 (1999).
\bibitem{Ber005912}
A. Bertoni, P. Bordone, R. Brunetti, C. Jacoboni, and S. Reggiani,
Phys. Rev. Lett. {\bf 84}, 5912 (2000).
\bibitem{Pri98141}
V. Privman, I. D. Vagner, and G. Kventsel, Phys. Lett. A {\bf 239}, 141 (1998).
\bibitem{Kit9707}
A.Yu. Kitaev, {\it Fault-tolerant quantum computation by anyons}, e-print quant-ph/9707021;
S. Lloyd, {\it Quantum computation with abelian anyons}, e-print quant-ph/0004010.

\bibitem{Bar95}
A. Barenco, D. Deustch, A. Ekert, and R. Jozsa, Phys. Rev. Lett. {\bf 74}, 4083 (1995).
\bibitem{San99}
G.D. Sanders, K.W. Kim, and W.C. Holton, Phys. Rev. A {\bf 60}, 4146 (1999).
\bibitem{Tan00}
T. Tanamoto, Phys. Rev. A {\bf 61}, 13813 (2000).
\bibitem{Li01}
X.Q. Li and Y. Arakawa, Phys. Rev. A {\bf 63}, 012302 (2001).
\bibitem{Bio00}
E. Biolatti, R.C. Iotti, P. Zanardi, and F. Rossi, Phys. Rev. Lett. {\bf 85}, 5647 (2000).
\bibitem{Qui99}
L. Quiroga and N.F. Johnson, Phys. Rev. Lett. {\bf 83}, 2270 (1999);
J.H. Reina, L. Quiroga, and N.F. Johnson, Phys. Rev. A {\bf 62}, 012305 (2000);
e-print quant-ph/0009035.
\bibitem{Sham01}
P. Chen, C. Piermarocchi, and L.J. Sham, {\it Control of Spin Dynamics of Excitons
in Nanodots for Quantum Operations}, e-print cond-mat/0102482.
\bibitem{She99}
M. S. Sherwin, A. Imamoglu, and T. Montroy, Phys. Rev. A {\bf 60}, 3508 (1999).
\bibitem{Ima99}
A. Imamoglu, D. D. Awschalom, G. Burkard, D. P. DiVincenzo,
D. Loss, M. Sherwin, and A. Small, Phys. Rev. Lett. {\bf 83}, 4204 (1999).
\bibitem{Bru00}
T. A. Brun, and H. Wang, Phys. Rev. A {\bf 61}, 032307 (2000).
\bibitem{Bro01}
K. R. Brown, D. A. Lidar, and K. B. Whaley,
{\it Quantum computing with quantum dots on quantum linear supports}, quant-ph/0105102.
\bibitem{Sham00}
G. Chen, N.H. Bonadeo, D.G. Steel, D. Gammon, D.S. Katzer, D. Park, L.J. Sham,
Science {\bf 289}, 1906 (2000).
\bibitem{Haw01}
M.Bayer, P. Hawrylak, K. Hinzer, S. Fafard, M. Korkusinski,
Z.R. Wasilewski, O. Stern, and A. Forchel, Science {\bf 291}, 451 (2001).
\bibitem{Gup01}
J.A. Gupta, R. Knobel, N. Samarth, and D.D. Awschalom,
Science {\bf 292}, 2458 (2001).
\bibitem{Lei95}
C. Leichtle, W.P. Schleich, I.S. Averbukh, and M. Shapiro,
Phys. Rev. Lett. {\bf 80}, 1418 (1995).
\bibitem{Wei9899}
T.C. Weinacht, J. Ahn, and P.H. Bucksbaum,
Phys. Rev. Lett. {\bf 80}, 5508 (1998);
Nature (London) {\bf 397}, 233 (1999).


\bibitem{Chi01}
A.M. Childs and I.L. Chuang, Phys. Rev. A {\bf 63}, 012306 (2001).
\bibitem{Efr96}
Al.L. Efros, M. Rosen, M. Kuno, M. Nirmal, D.J. Norris, and M. Bawendi,
Phys. Rev. B {\bf 54}, 4843 (1996).
\bibitem{Boc90}
U. Bockelman and G. Bastard,
Phys. Rev. B {\bf 42}, 8947 (1990).
\bibitem{Ben91}
H. Benisty, C.M. Sotomayor-Torres, and C. Weisbuch,
Phys. Rev. B {\bf 44}, 10945 (1991).

\bibitem{Li98}
X.Q. Li and Y. Arakawa,
Phys. Rev. B {\bf 57}, 12285 (1998).
\bibitem{Li99}
X.Q. Li, H. Nakayama, and Y. Arakawa,
Phys. Rev. B {\bf 59}, 5069 (1999).

\end{references}
\end{document}